\documentclass[prl,showpacs,groupedaddress,superscriptaddress,twocolumn,
10pt,nofootinbib,floatfix,preprintnumbers]{revtex4}
\usepackage{amssymb,amsmath,graphicx,xcolor,hyperref}

\begin{document}

\title{Flavor Structure of the Nucleon Sea from Lattice QCD}
\author{Huey-Wen Lin}
\email{hwlin@phys.washington.edu}
\affiliation{Department of Physics, University of Washington, Seattle, WA 98195-1560}
\author{Jiunn-Wei Chen}
\email{jwc@phys.ntu.edu.tw}
\affiliation{Department of Physics, National Center for Theoretical Sciences, and Leung
Center for Cosmology and Particle Astrophysics, National Taiwan University,
Taipei 10617, Taiwan}
\author{Saul D. Cohen}
\email{sdcohen@uw.edu}
\affiliation{Institute for Nuclear Theory, University of Washington, Seattle, WA
98195-1560}
\affiliation{Department of Physics, University of Washington, Seattle, WA 98195-1560}
\author{Xiangdong Ji}
\email{xji@umd.edu}
\affiliation{Maryland Center for Fundamental Physics, Department of Physics, University
of Maryland, College Park, Maryland 20742, USA}
\affiliation{INPAC, Department of Physics and Astronomy, Shanghai Jiao Tong University,
Shanghai, 200240, P. R. China}
\date{\today }
\pacs{
      12.38.Gc,       14.20.Dh,       14.65.Bt }
\preprint{{NT@UW-14-03, INT-PUB-14-002}}

\begin{abstract}
We present the first direct lattice calculation of the isovector sea-quark
distributions in the nucleon within the framework of the large-momentum
effective field theory proposed recently. We use $N_f=2+1+1$ HISQ lattice
gauge ensembles (generated by MILC Collaboration) and clover valence
fermions with pion mass 310~MeV. We establish the convergence of the result
as the nucleon momentum increases within the uncertainty of the calculation.
Although the lattice systematics are not yet fully under control, we obtain
some qualitative features of the flavor structure of the nucleon sea : $%
\overline{d}(x) > \overline{u}(x)$ leading to the violation of the Gottfried
sum rule; $\Delta \overline{u}(x) > \Delta \overline{d}(x)$ as indicated by
the STAR data at large and small leptonic pseudorapidity.
\end{abstract}

\maketitle



The proton has the quantum numbers of two up and one down quarks. The
fundamental theory for the proton structure, quantum chromodynamics (QCD),
predicts that in addition to these valence quarks, there is also a sea of
quark-antiquark pairs. The antiquarks in the proton can be probed in
high-energy scattering, particularly through the Drell-Yan process and
similar processes such as $W$ production. They can also be extracted through
semi-inclusive processes by tagging the fragmentations of the antiquarks. In
recent years, much progress has been made in understanding the flavor
structure of the nucleon sea (see Ref.~\cite{Chang:2014jba} for a recent
review), for the unpolarized sea in Drell-Yan~\cite%
{Adams:1995sh,Towell:2001nh} and for the polarized sea in semi-inclusive
deep-inelastic scattering~\cite{deFlorian:2009vb,Aschenauer:2013woa}.
Theoretical understanding of the nucleon sea has mostly been from nucleon
models~\cite{Speth:1996pz,Diakonov:1986yh}. Although the models provide a
qualitative physical understanding of the sea, they are not expected to make
reliable quantitative predictions. The only fundamental approach to nucleon
structure so far is lattice QCD. Unfortunately, the traditional lattice-QCD
approach does not allow one to compute the sea directly: one can only
calculate lower moments of the parton distributions, which involve quark as
well as antiquark contributions, making isolation of the antiquarks
difficult.

In a recent paper by one of us~\cite{Ji:2013dva}, a new approach to
calculating the full $x$-dependence of parton physics, such as the parton
distributions and other parton observables, has been proposed. The method is
based on the observation that, while in the rest frame of the nucleon,
parton physics corresponds to light-cone correlations, and the same physics
can be obtained through time-independent spatial correlations in the
infinite-momentum frame (IMF). For finite but large momenta feasible in
lattice simulations, a large-momentum effective field theory (LaMET) can be
used to relate Euclidean quasi-distributions to physical ones through a
factorization theorem~\cite{Ji:2014gla}.

In this paper, we report the first attempt to make a lattice calculation of
polarized and unpolarized quark distributions using the LaMET formalism. To
simplify the computation, we consider only the isovector $u-d$ combination
so that the disconnected diagrams do not contribute. We first compute the
Euclidean lattice quasi distribution $\tilde{q}_\text{lat}(x,\Lambda,P_z)$
at increasing nucleon momentum $P_z$ from 0.43 to 1.29~GeV and then extract
the physical light-cone distribution $q(x,\mu)$ by taking into account both
one-loop logarithmic and power corrections. The leading power correction
arises from the nucleon mass effect in the expansion with respect to $%
(M_N/4P_z)^2$. We observe the convergence of the result at the largest two
nucleon momenta, with small residual corrections coming from dynamical
higher-twist effects. Although the final result is not yet entirely physical
because of the large light-quark masses, coarse lattice spacing, and lack of
complete one-loop matching condition, we nonetheless find a qualitative
agreement with experimental data on unpolarized and polarized sea. This
demonstrates the feasibility of the approach and will motivate lattice-QCD
studies with improved systematics in the future.

For the quark distributions, the starting point is the momentum-dependent
nonlocal static correlation
\begin{align}
\tilde{q}(x,\Lambda,P_z) & = \int \frac{dz}{4\pi} e^{-izk} \times  \notag \\
& \left\langle \vec{P}\right\vert \bar{\psi}(z)\gamma_z e^{ig\int_0^z
A_z(z^\prime) dz^\prime} \psi(0) \left\vert \vec{P}\right\rangle ,
\label{eq:quark-dist}
\end{align}
where $x=k/P_z$, $\Lambda$ is an ultraviolet (UV) cut-off scale such as $1/a$
on a lattice with $a$ as lattice spacing, $\vec{P}$ is the momentum of the
nucleon moving in the $z$-direction. All fields and couplings are bare and
depend on $\Lambda$. When the nucleon momentum approaches infinity, the
quasi-distribution becomes the physical parton distribution when $\Lambda$
is fixed. At large but finite $P_z$, one has an effective field theory
expansion~\cite{Xiong:2013bka}
\begin{multline}
\tilde{q}(x,\Lambda,P_z) = \int \frac{dy}{\left\vert y\right\vert} Z\left(%
\frac{x}{y},\frac{\mu}{P_z}, \frac{\Lambda}{P_z}\right) q(y,\mu) \\
+ \mathcal{O}\left( \frac{\Lambda_\text{QCD}^2}{P_z^2}, \frac{M_N^2}{P_z^2}
\right) + \ldots .  \label{eq:z}
\end{multline}
where $\mu$ is the renormalization scale for the physical parton
distribution $q(y)$, usually in $\overline{\mathrm{MS}}$ scheme. The $Z$
function is a perturbation series in $\alpha_s$, depending among others on
the UV properties of the quasi-distribution. $Z$ has been calculated to
one-loop order in the transverse-momentum cutoff scheme, but is not yet
available in a lattice regularization. As observed in Ref.~\cite{Ma:2014jla}%
, the above relation can be inverted, since it is perturbative.

In this study, we use clover valence fermions on an ensemble of $24^3\times
64$ gauge configurations with $a \approx 0.12$~fm, box size $L\approx 3$~fm
and pion mass $M_\pi \approx 310$~MeV with $N_f=2+1+1$ flavors of highly
improved staggered quarks (HISQ) generated by MILC Collaboration~\cite%
{Bazavov:2012xda} and apply hypercubic (HYP) smearing~\cite%
{Hasenfratz:2001hp} to the gauge links. HYP smearing has been shown to
significantly improve the discretization effects on operators and shift
their corresponding renormalizations toward their tree-level values (near
unity for quark bilinear operators)~\cite{Bhattacharya:2013ehc}. We
calculate the quasi-distributions with long straight gauge-link products
between the quark and antiquark in the inserted current,
\begin{gather}
\tilde{q}_{\text{lat}}(x,\Lambda,P_z) = \int \frac{dz}{4\pi}
e^{-izk}h(z,\Lambda,P_z),  \notag \\
h(z,\Lambda,P_z) = \left\langle \vec{P}\right\vert \bar{\psi}(z) \gamma_z
\left( \prod_n U_z(n\hat{z})\right) \psi(0) \left\vert \vec{P}\right\rangle ,
\label{eq:qlatt}
\end{gather}
where $U_\mu$ is a discrete gauge link in the $\mu$ direction.

\begin{figure}[tbp]
\begin{center}
\includegraphics[width=0.45\textwidth]{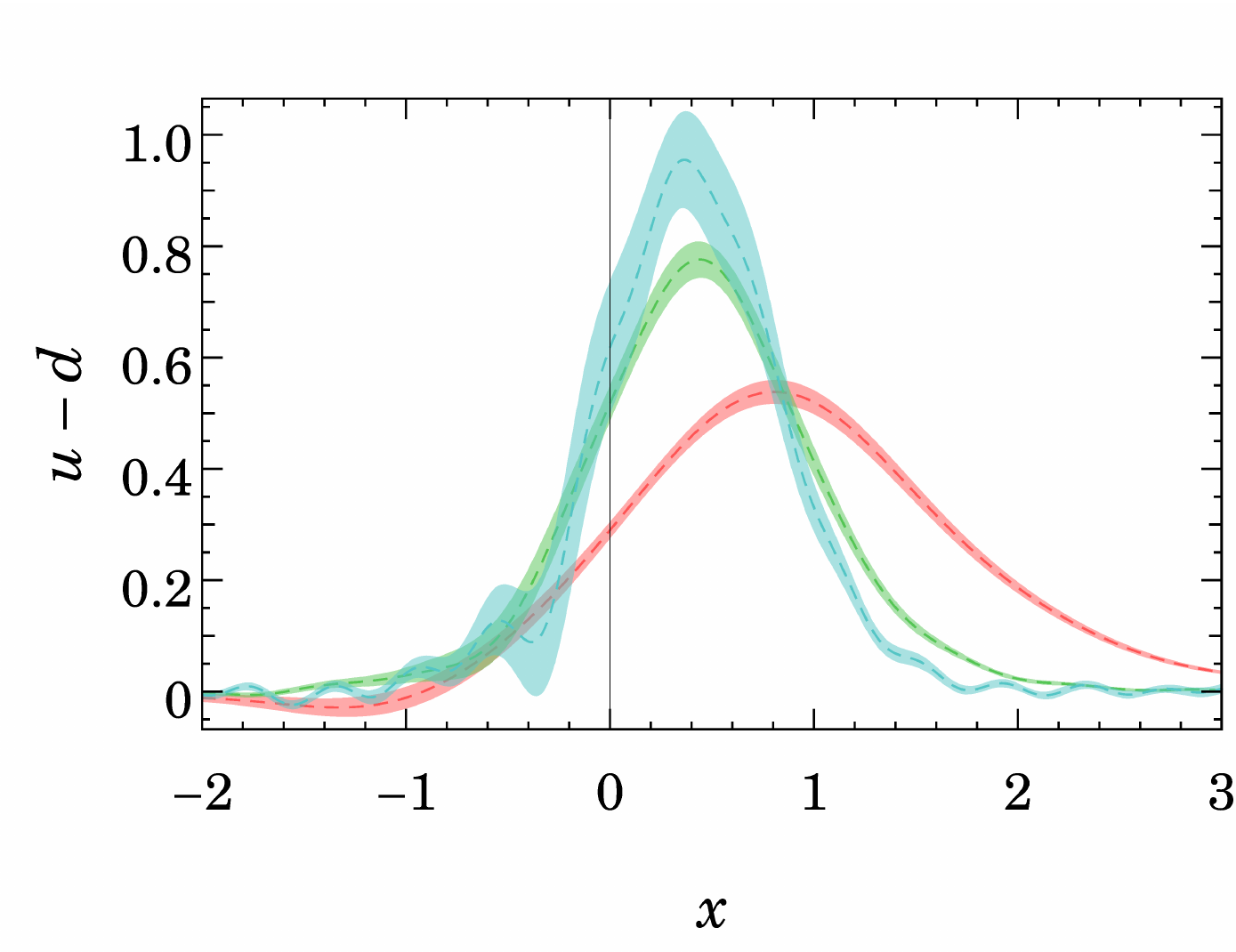}\\[%
0pt]
\end{center}
\caption{The isovector quark quasi-distribution $\tilde{u}(x)-\tilde{d}(x)$
as defined in Eq.~\protect\ref{eq:quark-dist} computed on a lattice with the
nucleon momentum $P_z$ (in units of $2\protect\pi /L$) $=$ 1 (red), 2
(green), 3 (cyan).}
\label{fig:quasi}
\end{figure}

We generate the results using 1383 measurements (among 461 lattice
configurations). We extract the matrix elements $h(z,\Lambda,P_z)$ for
various $z$ for our lattice setup with $P_z$ (in units of $2\pi/L$)1, 2, 3.
The statistical error becomes noticeably bigger as the nucleon momentum
becomes larger, as typically seen in lattice hadron calculations. The
correlation vanishes beyond about 1~fm, as is typical in nonperturbative
QCD. This is in strong contrast to the correlation in the light-cone
coordinates, as seen from the Fourier transformation of the parton
distribution in Feynman variable $x$, where the correlation length increases
with the nucleon momentum. In the present formalism, the small $x$ partons
arise from the spatial correlation of order 1~fm, whereas the valence parton
correlation is Lorentz contracted along the $z$ direction, as discussed in
Ref.~\cite{Ji:2014gla}.

We Fourier transform the $z$ coordinate into momentum $k$ to obtain the
quasi-distribution $\tilde{q}_\text{lat}(x,\mu,P_z)$, which is shown in Fig.~%
\ref{fig:quasi}.
It is quite striking that the peak at the lowest momentum is around $x=1$,
where the physical parton distribution vanishes. However, as the nucleon
momentum doubles, the peak shifts to $x\approx 0.5$ and the value of the
quasi-distribution at $x=1$ reduces to half that of the peak. At the highest
momentum, the peak is further shifted to $x\approx 0.4$ and the value at $%
x=1 $ is now about a third that of the peak. This is consistent with the
expectation that as momentum becomes asymptotically large, the
quasi-distribution becomes more similar to the physical parton distribution.
However, there is a limitation the size of the momentum available on the
lattice for nucleons. Therefore, LaMET must be used to extract the
asymptotic distribution from the finite-$P_z$ quasi-distributions. If we
account for all the corrections, any quasi-distribution at a reasonably
large $P_z$ should yield the same physical prediction.

To take into account the one-loop corrections, we use the $Z\left(\xi=\frac{x%
}{y} , \frac{\mu }{P_{z}},\frac{\Lambda}{P_z}\right) $ factor from Ref. \cite%
{Xiong:2013bka}. To make the computation easier, we use the inverted Eq.~\ref%
{eq:z} between the quasi- and physical distributions, expanded to linear
order in $\alpha_s$~\cite{Ma:2014jla}. We take the UV cutoff $\Lambda$ to be
the largest lattice momentum $\pi/a$, and $\overline{\mathrm{MS}}$ $\mu=2$
GeV, which sets the scale of the parton distribution. The choice of the
strong coupling is somewhat subtle.\footnote{%
In principle, one should use $6/(4 \pi\beta)$ on the lattice; however, it is
well known that this omits important tadpole contributions~\cite%
{Lepage:1992xa}. As a compromise, we take $\alpha_s=0.20\pm 0.04$, with the
central value determined by the prescription of Ref.~\cite{Lepage:1992xa}
and the uncertainty included as a part of the theoretical systematics.}
Again, the $Z$ factor from the cutoff scheme is correct to the leading
logarithm but not for the numerical constant. This is a compromise that we
make at the moment and will be rectified in the future.

At low nucleon momenta, the nucleon-mass corrections are as important as the
one-loop correction, if not more. Using the operator product expansion, the
nonlocal operator in Eq.~\ref{eq:quark-dist} can be expanded as $%
\sum_{n=1}^{\infty} C_n(z) O_n(0)$, where the tree-level Wilson coefficient $%
C_n(z)=\left(iz\right)^{n-1}/\left(n-1\right)! + \mathcal{O}(\alpha_s)$ and $%
O_n(0)= \bar{\psi}(0) \gamma^z \left(iD^z\right)^{n-1} \psi(0)$. The tensor $%
O_n$ is symmetric but not traceless, so it is a mixture of a twist-2 and
higher-twist operators with the matrix element
\begin{equation}
\left\langle \vec{P}\right\vert O_n(0)\left\vert \vec{P} \right\rangle = 2
a_n P_z^n K_n + \mathcal{O}(\Lambda_\text{QCD}^2/P_z^2)  \label{eq:Kn}
\end{equation}
entirely expressible in terms of $a_n=\int\!dx\,x^{n-1}q(x)$, the $n^\text{th%
}$ moment of the desired parton distribution, and $K_n = 1 +
\sum_{i=1}^{i_{\max}} C_{i}^{n-i} (M_N^2/4P_z^2)^i$ where $C$ is the
binomial function, and $i_{\max} = \frac{n-(n\bmod 2)}2$. The $\mathcal{O}%
(\Lambda_\text{QCD}^2/P_z^2)$ term is dynamical higher-twist correction. As
one can see, the actual nucleon-mass correction parameter is $M_N^2/4P_z^2$.

\begin{figure}[tbp]
\begin{center}
\includegraphics[width=0.45\textwidth]{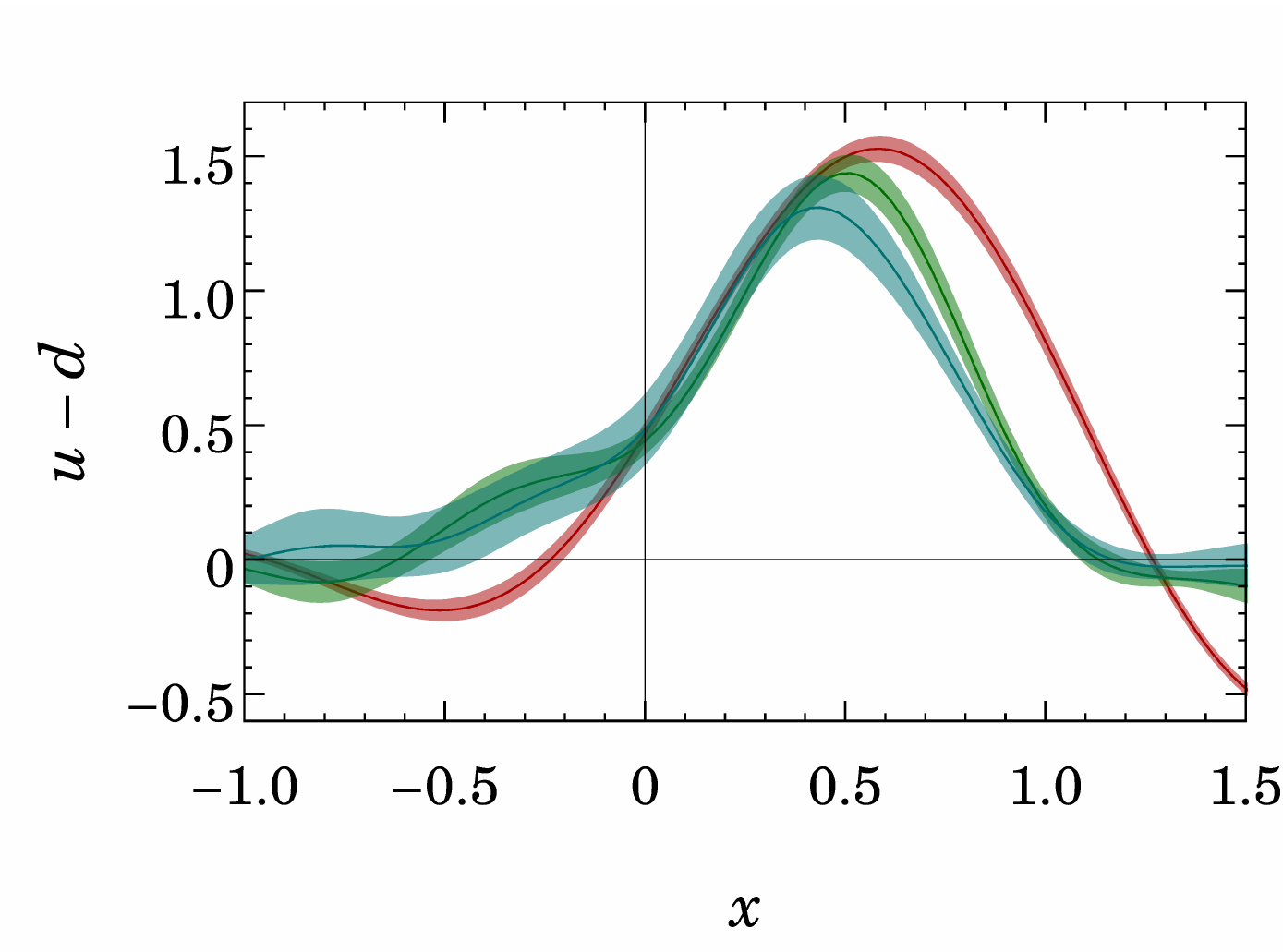}
\end{center}
\caption{The physical quark distribution $u(x)-d(x)$ extracted from Fig.~%
\protect\ref{fig:quasi} after making $M_N^n/P_z^n$ corrections and one-loop
corrections. The red, green and cyan bands correspond to $P_z \in \{1,2,3\}%
\frac{2\protect\pi}{L}$. The two higher-momentum distributions are now
almost identical.}
\label{fig:quark-dist1}
\end{figure}

After one-loop and nucleon-mass corrections, the resulting distributions are
shown in Fig.~\ref{fig:quark-dist1}. For the nuclear momenta under
consideration, both types of correction are important. As one can see, the
corrected distributions have much reduced $P_{z}$ dependence, particularly
for the two largest momenta. This suggests that the corrections to the
quasi-distributions will generate a $P_{z}$-independent physical
distribution. The remaining small difference between the two large-momenta
results could be due to the dynamical higher-twist corrections $\mathcal{O}%
(\Lambda _{\text{QCD}}^{2}/P_{z}^{2})$, which is expected to be smaller than
the nucleon-mass effect. As for the lowest nucleon momentum (430~MeV)
result, the LaMET expansion might not be very effective, although the peak
after corrections has been shifted to near $0.8$.

\begin{figure}[tbp]
\begin{center}
\includegraphics[width=0.45\textwidth]{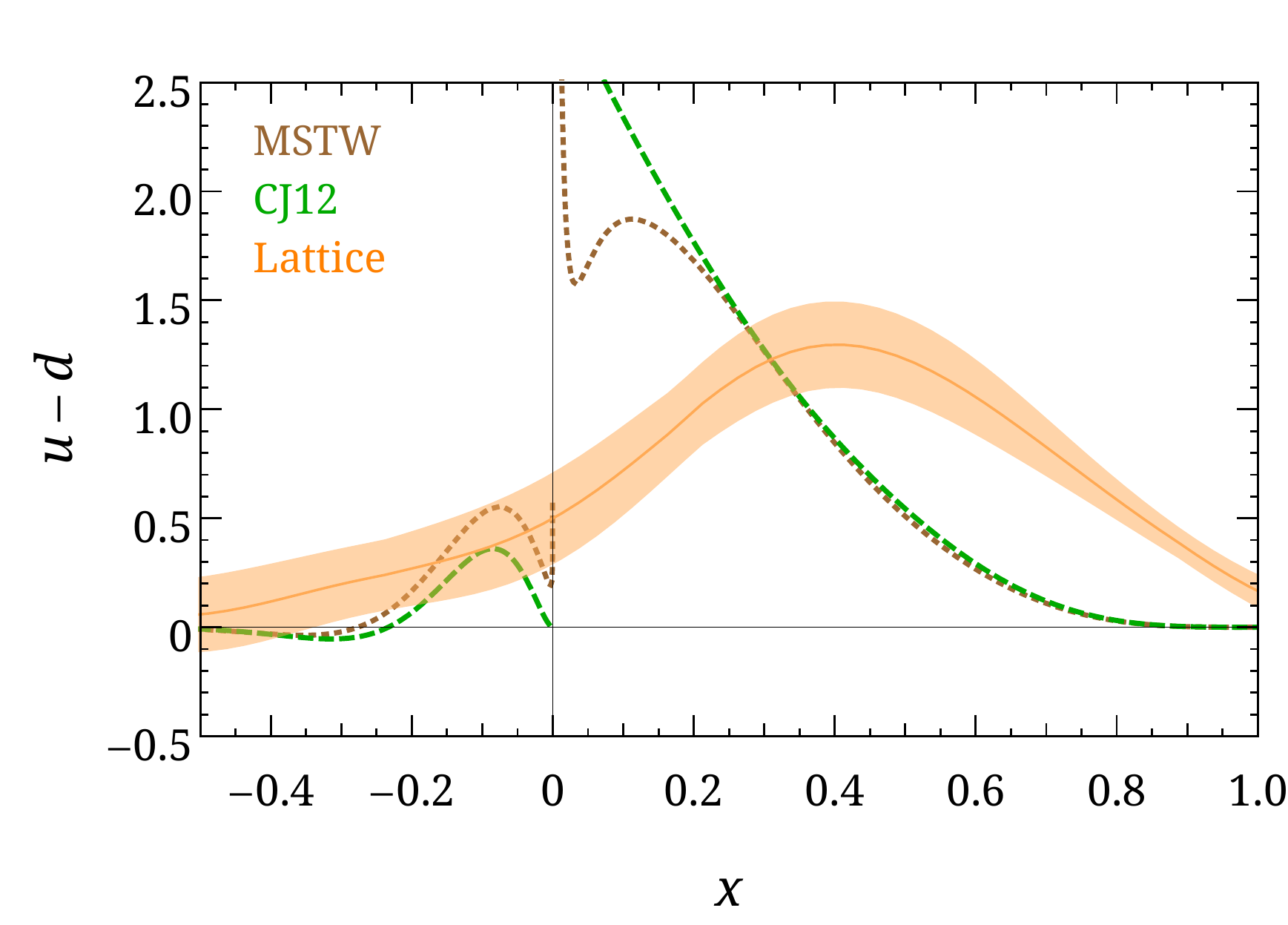}
\end{center}
\caption{The unpolarized isovector quark distribution $u(x)-d(x)$ computed
on the lattice after extrapolation in $P_{z}$ is shown as the purple band,
compared with the global analyses by MSTW~\protect\cite{Martin:2009iq}
(brown dotted line), and CTEQ-JLab (CJ12, green dashed line)~\protect\cite%
{Owens:2012bv} with medium nuclear correction near $(1.3\text{ GeV})^{2}$.
The negative $x$ region is the sea quark distribution with $\overline{q}%
(x)=-q(-x)$. The lattice uncertainty band in the plot reflects the 68\% C.L.
The global fit uncertainty is not shown in the figure.}
\label{fig:quark-dist2}
\end{figure}

Finally, we find a $P_z$-independent distribution by taking into account the
$\mathcal{O}(\Lambda_\text{QCD}^2/P_z^2)$ correction by extrapolating using
the form $a + b /P_z^2$. The final unpolarized distribution $u(x)-d(x)$ is
shown in Fig.~\ref{fig:quark-dist2}. The distribution for the $|x| > 1$
region is within 2 sigma of zero; thus, we recover the correct support for
the physical distribution within error.

%

Our result cannot be directly compared with the experimental data because
other lattice systematics are not yet under control. To obtain the physical
parton distributions, we need to make a number of improvements, including
reducing the quark masses to physical ones, increasing the number of
configurations to reduce statistical errors, using finer lattice spacing to
accommodate larger boosted momenta and improve the resolution, and using
larger lattice volumes to access smaller $x$. Nonetheless, we hope that the
present results do provide some insight into the qualitative features of the
parton physics.

Also shown in Fig.~\ref{fig:quark-dist2} are the parton distributions from
the global analyses by CTEQ-JLab (CJ12)~\cite{Owens:2012bv} and NLO MSTW08~%
\cite{Martin:2009iq} at $\mu \approx 1.3$~GeV. Note that the lattice results
are not yet close to the physical pion mass, and the comparison with global
analysis here is mainly to demonstrate the magnitude of the quantity, rather
than to attempt to make a detailed comparison. The distribution in the mid-$%
x $ region between 2 difference scales 1.3 GeV and 2 GeV is small, comparing
with the changes we anticipate in the future calculation at physical pion
mass. In Fig.~\ref{fig:quark-dist2}, the lattice distribution weighs more at
larger $\left\vert x\right\vert $. And since the total $u-d$ quarks are
conserved, a reduction in small $\left\vert x\right\vert $ means an increase
in larger $\left\vert x\right\vert $. This is also consistent with that the
lattice first-moment of the momentum fraction ($\langle x\rangle _{u-d}$)
and helicity ($\langle x\rangle _{\Delta u-\Delta d}$) above pion mass
250~MeV is roughly double the integrated values derived from global analyses~%
\cite{Lin:2009qs,Alexandrou:2013cda}. It would be very interesting to
observe how the distribution changes when the lattice systematics improve.

The sea-quark distribution can be read from the negative-$x$ contribution: $%
\overline{q}(x)=-q(-x)$. Our result favors a large asymmetry in the
distributions of sea up and down antiquarks in the nucleon. There is a
violation of the Gottfried sum rule with $\int_{0}^{\infty }dx\,(\overline{u}%
(x)-\overline{d}(x))=0.14(5)$, which was first observed by New Muon
Collaboration (NMC) through the cross-section ratio for deep inelastic
scattering of muons from hydrogen and deuterium~\cite{Arneodo:1994sh}, and
later confirmed by other experiments using different processes, such as
Drell-Yan at E665~\cite{Adams:1995sh} and E866/NuSea~\cite{Towell:2001nh}.
This is the first time we can demonstrate this directly from lattice QCD.
Our result is close to the experimental one obtained by NMC in their DIS
measurement, $0.147(39)$ at $Q^{2}=4\text{ GeV}^{2}$ and by HERMES in their
semi-inclusive DIS (SIDIS) result, $0.16(3)$ at $Q^{2}=2.3\text{ GeV}^{2}$~%
\cite{Ackerstaff:1998sr}.


The study of the isovector helicity distribution follows the same procedure
with $\gamma_z$ in Eq.~\ref{eq:qlatt} replaced by $\gamma_z\gamma_5$.
Our result for antiquark helicity favors more polarized up quark than down
flavor, while the total polarized sea asymmetry estimated by DSSV09 is
consistent with zero within 2~sigma. We see a bigger polarized sea
asymmetry, $\Delta \overline{u}-\Delta \overline{d}=0.24(6)$, than the
unpolarized case, as predicted in the large-$N_c$ theory. We also see more
weight distributed near the $x=1$ regions, which could shift as we lower the
light-quark masses in the future. In the near term, experiments in
longitudinal single-spin asymmetry and parity-violating $W$ production at
RHIC might shed more light on the polarized sea distribution~\cite%
{Aschenauer:2013woa}.

To summarize, we have presented a direct lattice-QCD calculation of the $x$
dependence of parton distribution functions. By doing calculations with a
large-momentum nucleon, we are able to connect light-cone quantities to
lattice-QCD nonlocal but time-independent matrix elements. Since the largest
attainable momentum is limited, we correct for the sizable finite-momentum
dependence systematically. Our final result shows very encouraging signal
for the isospin sea asymmetry in the unpolarized quark and helicity
distributions. It is first time these are directly calculated with a
first-principles nonperturbative QCD approach. There is no fundamental
difficulty in performing the calculation at the physical pion mass, and
improving the statistical error to a level where we can compare with
experiments quantitatively.


\begin{acknowledgments}
The LQCD calculations were performed using the Chroma software suite~\cite%
{Edwards:2004sx} on Hyak clusters at the University of Washington managed by
UW Information Technology, using hardware awarded by NSF grant PHY-09227700.
We thank MILC collaboration for sharing the lattices used to perform this
study. The work of HWL and SDC is supported by the DOE grant
DE-FG02-97ER4014 and DE-FG02-00ER41132. JWC is supported in part by the NSC,
NTU-CTS, and the NTU-CASTS of R.O.C. The work of XJ is partially supported
by the U.~S. Department of Energy via grants DE-FG02-93ER-40762, and a grant
(No. 11DZ 2260700) from the Office of Science and Technology in Shanghai
Municipal Government, and by National Science Foundation of China
(No.~11175114). HWL would like to thank the Shanghai Jiao Tong University
for their hospitality during the development of the ideas presented in this
work.
\end{acknowledgments}


\bibliographystyle{plain}
\bibliography{parton_sea}

\begin{thebibliography}{10}

\bibitem{Ackerstaff:1998sr}
K.~Ackerstaff et~al.
\newblock {The Flavor asymmetry of the light quark sea from semiinclusive deep
  inelastic scattering}.
\newblock {\em Phys.Rev.Lett.}, 81:5519--5523, 1998.

\bibitem{Adams:1995sh}
M.R. Adams et~al.
\newblock {Extraction of the ratio $F2_n / F2_p$ from muon - deuteron and muon
  - proton scattering at small $x$ and Q**2}.
\newblock {\em Phys.Rev.Lett.}, 75:1466--1470, 1995.

\bibitem{Alexandrou:2013cda}
C.~Alexandrou, M.~Constantinou, V.~Drach, K.~Hatziyiannakou, K.~Jansen, et~al.
\newblock {Nucleon Structure using lattice QCD}.
\newblock {\em Nuovo Cim.}, C036(05):111--120, 2013.

\bibitem{Arneodo:1994sh}
M.~Arneodo et~al.
\newblock {A Reevaluation of the Gottfried sum}.
\newblock {\em Phys.Rev.}, D50:1--3, 1994.

\bibitem{Aschenauer:2013woa}
E.C. Aschenauer, A.~Bazilevsky, K.~Boyle, K.O. Eyser, R.~Fatemi, et~al.
\newblock {The RHIC Spin Program: Achievements and Future Opportunities}.
\newblock 2013.

\bibitem{Bazavov:2012xda}
A.~Bazavov et~al.
\newblock {Lattice QCD ensembles with four flavors of highly improved staggered
  quarks}.
\newblock {\em Phys.Rev.}, D87(5):054505, 2013.

\bibitem{Bhattacharya:2013ehc}
Tanmoy Bhattacharya, Saul~D. Cohen, Rajan Gupta, Anosh Joseph, Huey-Wen Lin,
  et~al.
\newblock {Nucleon Charges and Electromagnetic Form Factors from 2+1+1-Flavor
  Lattice QCD}.
\newblock {\em Phys.Rev.}, D89:094502, 2014.

\bibitem{Chang:2014jba}
Wen-Chen Chang and Jen-Chieh Peng.
\newblock {Flavor Structure of the Nucleon Sea}.
\newblock 2014.

\bibitem{deFlorian:2009vb}
Daniel de~Florian, Rodolfo Sassot, Marco Stratmann, and Werner Vogelsang.
\newblock {Extraction of Spin-Dependent Parton Densities and Their
  Uncertainties}.
\newblock {\em Phys.Rev.}, D80:034030, 2009.

\bibitem{Diakonov:1986yh}
Dmitri Diakonov and V.~Yu. Petrov.
\newblock {Chiral Theory of Nucleons}.
\newblock {\em JETP Lett.}, 43:75--77, 1986.

\bibitem{Edwards:2004sx}
Robert~G. Edwards and Balint Joo.
\newblock {The Chroma software system for lattice QCD}.
\newblock {\em Nucl.Phys.Proc.Suppl.}, 140:832, 2005.

\bibitem{Hasenfratz:2001hp}
Anna Hasenfratz and Francesco Knechtli.
\newblock {Flavor symmetry and the static potential with hypercubic blocking}.
\newblock {\em Phys.Rev.}, D64:034504, 2001.

\bibitem{Ji:2013dva}
Xiangdong Ji.
\newblock {Parton Physics on Euclidean Lattice}.
\newblock {\em Phys.Rev.Lett.}, 110:262002, 2013.

\bibitem{Ji:2014gla}
Xiangdong Ji.
\newblock {Parton Physics from Large-Momentum Effective Field Theory}.
\newblock {\em Sci.China Phys.Mech.Astron.}, 57:1407--1412, 2014.

\bibitem{Lepage:1992xa}
G.~Peter Lepage and Paul~B. Mackenzie.
\newblock {On the viability of lattice perturbation theory}.
\newblock {\em Phys.Rev.}, D48:2250--2264, 1993.

\bibitem{Lin:2009qs}
Huey-Wen Lin.
\newblock {A Review of Nucleon Spin Calculations in Lattice QCD}.
\newblock {\em AIP Conf.Proc.}, 1149:552--557, 2009.

\bibitem{Ma:2014jla}
Yan-Qing Ma and Jian-Wei Qiu.
\newblock {Extracting Parton Distribution Functions from Lattice QCD
  Calculations}.
\newblock 2014.

\bibitem{Martin:2009iq}
A.D. Martin, W.J. Stirling, R.S. Thorne, and G.~Watt.
\newblock {Parton distributions for the LHC}.
\newblock {\em Eur.Phys.J.}, C63:189--285, 2009.

\bibitem{Owens:2012bv}
J.F. Owens, A.~Accardi, and W.~Melnitchouk.
\newblock {Global parton distributions with nuclear and finite-$Q^2$
  corrections}.
\newblock {\em Phys.Rev.}, D87:094012, 2013.

\bibitem{Speth:1996pz}
J.~Speth and Anthony~William Thomas.
\newblock {Mesonic contributions to the spin and flavor structure of the
  nucleon}.
\newblock {\em Adv.Nucl.Phys.}, 24:83--149, 1997.

\bibitem{Towell:2001nh}
R.S. Towell et~al.
\newblock {Improved measurement of the anti-d / anti-u asymmetry in the nucleon
  sea}.
\newblock {\em Phys.Rev.}, D64:052002, 2001.

\bibitem{Xiong:2013bka}
Xiaonu Xiong, Xiangdong Ji, Jian-Hui Zhang, and Yong Zhao.
\newblock {One-Loop Matching for Parton Distributions: Non-Singlet Case}.
\newblock {\em Phys.Rev.}, D90:014051, 2014.

\end{thebibliography}

\end{document}